\documentclass[sigconf,natbib=true,anonymous=false]{acmart}

 \setcopyright{acmcopyright}

\copyrightyear{2024}
\acmYear{2024}
\setcopyright{rightsretained}
\acmConference[SIGIR '24]{Proceedings of the 47th International ACM SIGIR Conference on Research and Development in Information Retrieval}{July 14--18, 2024}{Washington, DC, USA}
\acmBooktitle{Proceedings of the 47th International ACM SIGIR Conference on Research and Development in Information Retrieval (SIGIR '24), July 14--18, 2024, Washington, DC, USA}
\acmDOI{10.1145/3626772.3657964}
\acmISBN{979-8-4007-0431-4/24/07}

\usepackage{multirow}

\usepackage{enumitem}
\usepackage{todonotes}
\usepackage{xcolor}
\DeclareTextFontCommand{\judgment}{\bfseries\itshape}

\newcommand{\name}{PLAID SHIRTTT}

\begin{document}
\title{\name\ for Large-Scale Streaming Dense Retrieval}

\author{Dawn Lawrie}
\affiliation{%
  \institution{HLTCOE, Johns Hopkins University}
  \city{Baltimore}
  \state{Maryland}
  \country{USA}
}
\email{lawrie@jhu.edu}

\author{Efsun Kayi}
\affiliation{%
  \institution{HLTCOE, Johns Hopkins University}
  \city{Baltimore}
  \state{Maryland}
  \country{USA}
}
\email{ekayi1@jhu.edu}

\author{Eugene Yang}
\affiliation{%
  \institution{HLTCOE, Johns Hopkins University}
  \city{Baltimore}
  \state{Maryland}
  \country{USA}
}
\email{eugene.yang@jhu.edu}

\author{James Mayfield}
\affiliation{%
  \institution{HLTCOE, Johns Hopkins University}
  \city{Baltimore}
  \state{Maryland}
  \country{USA}
}
\email{mayfield@jhu.edu}

\author{Douglas W. Oard}
\affiliation{
    \institution{University of Maryland}
    \city{College Park}
    \state{Maryland}
    \country{USA}
}
\email{oard@umd.edu}

\begin{abstract}
PLAID, an efficient implementation of the ColBERT late interaction bi-encoder using pretrained language models for ranking,
consistently achieves state-of-the-art performance in monolingual, cross-language, and multilingual retrieval.  
PLAID differs from ColBERT by assigning terms to clusters and representing those terms as cluster centroids plus compressed residual vectors. 
While PLAID is effective in batch experiments,
its performance degrades in streaming settings where documents arrive over time
because representations of new tokens may be poorly modeled by the earlier tokens used to select cluster centroids. 
PLAID Streaming Hierarchical Indexing that Runs on Terabytes of Temporal Text (\name) addresses this concern using multi-phase incremental indexing based on hierarchical sharding. 
Experiments on ClueWeb09 and the multilingual NeuCLIR collection demonstrate the effectiveness of this approach both for the largest collection indexed to date by the ColBERT architecture and in the multilingual setting, respectively.

\end{abstract}




\maketitle              %

\section{Introduction}

Ranked retrieval using pretrained language models (PLMs) has shown great promise
on research test collections~\cite{dloverview2022}.
Two main architectures have emerged, cross-encoders and bi-encoders~\cite{lin2022pretrained}.
Cross-encoders are generally used as rerankers
because they must
process the query and the document passages together. 
Often a lexical matcher like BM25~\cite{manning:Intro2IR} is used to retrieve the documents to be reranked,
meaning that while a cross-encoder can match queries and documents semantically,
it will never be presented with documents that lack an exact query string match.\footnote{Of course, the query for a lexical matcher can result from automatic query ,rewriting and thus need not be the same as the query presented to the cross-encoder.}

A bi-encoder, by contrast, encodes document passages separately from the query,
enabling encoding in an offline indexing phase using GPUs. 
At query time, only the query needs to be encoded,
which, given its short length,
can be done quickly using a CPU.
A bi-encoder's dense representations can rank documents that are semantically similar to the query, even without exact string matches.
Moreover, if the bi-encoder's token representations are built from a multilingual Pretrained Language Model,
documents in languages other than that of the query can also be ranked.
This architecture could enable multilingual retrieval on the web.
Bi-encoders offer the promise of flexible and effective first-stage ranking.

Because of encoder limitations, bi-encoders normally break documents into passages;
a useful heuristic is to use the highest passage score as the document score~\cite{dai2019deeper}.
A bi-encoder can encode a passage using one or many vectors.
The single vector approach, for which the present state of the art is Contriever~\cite{izacard2021unsupervised},
is efficient and effective in monolingual tasks. 
A query is also represented as a single vector.
Passages are ranked by comparing the query vector to the passage vector.
But in multilingual tasks such as Cross-Language Information Retrieval (CLIR),
it is outperformed by ColBERT~\cite{colbert}
(our focus in this paper),
where each token is represented by a dense vector.
At search time each query token is represented as a vector and passages are ranked based on the passage tokens that are closest to each query token.
ColBERT is currently the state of the art for full-collection (i.e., end-to-end) CLIR~\cite{neucliroverview2022} and Multilingual Information Retrieval (MLIR)~\cite{lawrie2023neural}.
PLAID~\cite{plaid}, a space-efficient implementation of ColBERT,
is thus an obvious architecture to consider for a high-volume multilingual document stream. 

PLAID was designed for batch settings, because it needs access to all (or nearly all) the documents at the start of indexing. 
That is impractical in streaming settings, where documents are introduced over time.
This paper proposes \name\ 
(PLAID Streaming Hierarchical Indexing that Runs on Terabytes of Temporal Text).
The key blocker to incremental indexing for streaming in the PLAID architecture is its reliance on cluster centroids 
for term representation.
As the vocabulary in the new documents moves away from that in the documents from which the cluster centroids were built,
performance degrades precipitously. 
This paper proposes hierarchical sharding to adapt PLAID to a streaming setting.
Its main contributions include the architecture (\name),\footnote{Code available at \url{https://github.com/hltcoe/colbert-x}.}
evaluation that demonstrates its effectiveness in both monolingual and multilingual settings,
and the first known application of a ColBERT variant to a terabyte collection (ClueWeb09~\cite{callan2009clueweb09}).

\begin{table}[t]
  \caption{Effectiveness of centroids built at different points along the Chinese NeuCLIR stream.}
  \label{tab:freq}
  \centering

\begin{tabular}{l|ccc|c}
\toprule
Stream Percent &   5\% &  75\% &  90\% & 100\% \\
\midrule
nDCG@20        & 0.047 & 0.336 & 0.384 & \textbf{0.440} \\
R@1000         & 0.301 & 0.540 & 0.655 & \textbf{0.795} \\
\bottomrule
\end{tabular}

\end{table}

\section{Background}
\label{sec:background}

PLAID~\cite{plaid} and ColBERTv2~\cite{colbertv2} addressed one of the main disadvantages of ColBERT~\cite{colbert}--the large index size %
(3 GB of text requires a multi-vector retrieval index over 170 GB ).
This space is mainly consumed by dense term vectors.
PLAID reduced index storage by 
approximating each term vector
as the combination of one of a few canonical vectors
with a residual.
Canonical vectors are created by K-means clustering of term vectors drawn from a sampled set of documents,
then selecting the centroid of each cluster.
During indexing, the centroid nearest each document term is identified
and a residual is computed to represent the distance between the centroid and the term vector.
Only the cluster id and the residual is stored.
The fewer bits used to represent the residual, the less storage required for the index
but also the more lossy the compression.
At retrieval time, document scores are the sum of the distance between each query term its nearest document term;
closer document terms get higher scores. 

When using PLAID in a streaming environment,
we would like fixed cluster centroids.
Then all newly arriving documents could be indexed
by finding each term's nearest centroid from this fixed set and computing their residuals.
Table~\ref{tab:freq} shows that clustering once early in the stream does not work well.
For instance, if centroids are generated from the first 5\% of the chronologically ordered Chinese documents in the NeuCLIR test collection (see Section~\ref{sec:exp}),
nDCG@20 drops precipitously from 0.440 to 0.047.
Even waiting until 75\% of the stream has arrived still yields rather poor results (nDCG@20 0.336).
The reason for this degradation is that language use changes over time.
The NeuCLIR Collection spans 2016 to 2021,
so the centroids created early in the stream do not, for example, have a cluster for COVID,
and the nearest centroid cannot represent that concept well
(because compressed residuals are lossy).
One solution is to periodically re-compute cluster centroids.
Doing so results in new centroids,
thus requiring all prior documents to be re-indexed if retrospective search is to be supported.

Sharding is a common approach when collection size exceeds the capacity of a single index server~\cite{brown1994fast}.
Streaming adds the additional complexity of non-stationary statistics;
from the first TREC routing task it was understood that in streaming tasks,
collection statistics must be modeled using prior data~\cite{buckley1995automatic}.
Many streaming evaluations, such as the TREC Filtering~\cite{robertson2002trec} and Knowledge Base Acceleration (KBA)~\cite{frank2014evaluating} tracks,
focus on making yes/no decisions as documents arrive;
in contrast, our focus is on optimally supporting ranked retrieval over the full collection through the current time.  
We therefore introduce \name,
which uses hierarchical sharding to balance indexing efficiency
with the efficiency and effectiveness of full-collection ranked retrieval in streaming settings.

\section{\name}

\begin{figure}[t]
    \centering
    \includegraphics[width=\linewidth]{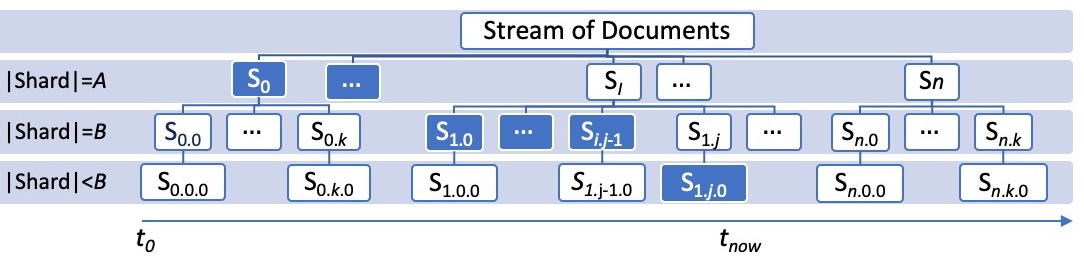}
    \caption{The sharded stream. Large shards of size $A$ are further partitioned into shards of $B$ documents. At any point in the stream, the last shard is incomplete. The shaded boxes represent the shards that make up the index at $t_{now}$.} 
    \label{fig:index}
\end{figure}

PLAID's cluster centroids are themselves derived from an underlying PLM such as BERT~\cite{bert};
ultimately the PLM determines document scores.
Our approach to indexing streaming documents accumulates documents into shards.
Each {\it{shard model}} is the set of dense vectors representing the cluster centroids
of the document terms in that shard. 
There is tension between creating as large a shard as possible
(to minimize the number of shards that must be searched)
and creating shard models for newly arrived documents as soon as possible 
(since a shard model based on earlier documents can be suboptimal).
\name\ addresses this tension by indexing each document a fixed, small number of times
to balance document ranking effectiveness
against the efficiency costs of re-indexing and of having too many shards.
Figure~\ref{fig:index} illustrates re-indexing three times.
The first time a document $d$ is indexed, a prior shard model is used.
Once a sufficient number $B$ of documents has arrived
a shard model containing document $d$ is created,
and document $d$ is re-indexed.
Once $k \times B=A$ documents have arrived, the final shard model containing document $d$ is created,
and document $d$ is once again re-indexed.
The edge case of the start of the stream is handled by running ColBERT V1
(without centroids or residuals)
until enough documents have arrived for an initial shard model. 

\begin{table}[]
\caption{Collection Statistics}\label{tab:collection-stats}
\centering
\begin{tabular}{l|cc|cc}
\toprule
Collection                &     \multicolumn{2}{c|}{ClueWeb09}  &  \multicolumn{2}{c}{NeuCLIR} \\
TREC Tracks                &  Web 09-12  & 2012 &     2022 &    2023 \\
\midrule
\# of Docs      &     \multicolumn{2}{c|}{504M}    &  \multicolumn{2}{c}{10.00M} \\
\# of Passages  &        \multicolumn{2}{c|}{31B} &  \multicolumn{2}{c}{58.9M} \\
\midrule
\# of Topics    &          200 &              50 &       41 &      65 \\
\midrule
\# Shards & \multicolumn{2}{c|}{108}    &  \multicolumn{2}{c}{21} \\
Sizes($A$/$B$) & \multicolumn{2}{c|}{5M/500K}    &  \multicolumn{2}{c}{500K/100K} \\
\bottomrule
\end{tabular}
\end{table}

Hierarchical sharding limits the number of shards that need to be searched at any time.
Each shard requires a CPU with sufficient memory to load the inverted cluster centroid index to support interactive search.
For instance, searching one hundred shards requires one hundred CPUs, each with access to hundreds of gigabytes of memory. 
Thus, the optimal settings of $A$ and $B$ depend on the rate of the stream and the resources available to support interactive search.
It may be advantageous to introduce more layers of the hierarchy to maintain a reasonable number of shards. Since each layer of the hierarchy involves re-indexing the documents seen so far using GPUs, this cost must also be considered when setting shard sizes.
Finally, PLAID requires at least 2000 passages,\footnote{\url{https://github.com/stanford-futuredata/ColBERT/issues/181\#issuecomment-1613956350}}
(each document may comprise more than one passage).
We choose $A$ to create the largest shards our hardware will accommodate, 
and $B$ to balance the number of shards with the arrival rate
(with larger $B$, documents will be represented by older shard models longer). 
We assume that when implementing \name\ that both the data-rates and how long documents must remain in the index are known. 

At search time, each active shard is searched using PLAID's standard approach.
Because all document scores are based on the same PLM
(differing only in the effect of the lossy residuals),
they are comparable and can be merged.
Efficient implementation will search shards in parallel,
merging results only for the highest ranked documents from each shard. %

\section{Experiment Design}
\label{sec:exp}

Our experiments seek to answer three questions:
(1) Is \name\ effective for large collections that arrive as a stream?
(2) How efficient is \name? and
(3)  How effective is \name\ for MLIR?
Table~\ref{tab:collection-stats} provides statistics for our two test collections. 
We chose ClueWeb09~\cite{callan2009clueweb09} because its size matches our aspirations to work at large scale;
retrieval evaluation is supported by extant relevance judgments for a large number of topics,
and the documents can be date-ordered.
This collection was used in four years of the TREC Web Track,
where web-style topics of the era were developed.
We report results over all 200 topics~\cite{webtrack2009, webtrack2010, webtrack2011, webtrack2012}.
The query field of a topic was often vague;
however, because dense retrieval benefits from queries with more context,
we combined the query and description fields to form our queries.
The reusability of this collection is questionable~\cite{sakai2013unreusability}
when using evaluation measures that treat unjudged documents as not relevant.
We follow Sakai~\cite{sakai2013unreusability} in using the compressed-list metrics MAP$'$ and nDCG$'$@20
that treat unjudged documents as not retrieved.

ClueWeb emphasizes scale, whereas NeuCLIR emphasizes multilinguality.
With ten million documents from the news subset of CommonCrawl,
NeuCLIR is presently the largest available test collection for MLIR; 
It is tiny though relative to ClueWeb, and thus has relatively small shards.
The NeuCLIR MLIR task is to retrieve relevant documents from any of the collection's three languages
(Chinese, Persian, Russian). 
For queries, we follow the majority of TREC 2022 and 2023 NeuCLIR track participants~\cite{neuclir2022,  neuclir2023}
in using the title and description fields of each of the 41 and 65 topics as queries, respectively.
We report nDCG@20, the primary measure reported by the TREC 2022 and 2023 NeuCLIR tracks, as well as other measures.

For each collection, we simulate a stream by ordering the documents by date,
arbitrarily ordering documents with the same timestamp.\footnote{\url{https://huggingface.co/datasets/hltcoe/plaid-shirttt-doc-date}} %
We extract text from  web pages using the {\texttt{newspaper}}\footnote{\url{https://pypi.org/project/newspaper3k/}} Python package,
which also associates a date with each article.\footnote{Persian dates before 1900 are converted to the Gregorian calendar.}
If {\texttt{newspaper}} cannot identify a date,
we examine the {\texttt{warc}} header for dates,
preferring the {\texttt{last-modified}} field over the {\texttt{date}} field.
If this too fails, we use the article crawl date.
While NeuCLIR was released with text extracted by {\texttt{newspaper}},
ClueWeb09 was released as raw web pages.
About three million documents that fail processing or lack both a title and content once processed are not indexed.

Our ClueWeb09 runs use the ColBERTv2 checkpoint fine-tuned from a BERT~\cite{bert} Base model,
as released by the ColBERT authors~\cite{colbertv2}.
For NeuCLIR, we use the ColBERT-X model,
fine-tuned with multilingual Translate-Train~\cite{lawrie2023neural} from XLM-RoBERTa Large model~\cite{xlmr}
on English to Chinese, Persian, and Russian. 
We store one bit for each residual vector dimension.
At search time, we retrieve the top fifty (Web topics) or one thousand (NeuCLIR topics) passages from each shard
and aggregate passage scores with MaxP~\cite{dai2019deeper} to form document scores.  
Table~\ref{tab:collection-stats} shows how we set the values of $A$ and $B$ for large and small shards respectively.
The sizes for ClueWeb reflect our recommendation to balance hardware limits against number of shards;
NeuCLIR sizes were selected to force creation of a non-trivial number of shards.

\section{Results}

Our first research question asks whether \name\ is effective for a large streaming collection. 
We use ClueWeb09 for these experiments. 
Results in Table~\ref{tab:clueweb09} report performance on topics from 2012,
alongside the best run from that year's TREC Web track for reference, uogTrA44xu~\cite{webtrack2012}.
These runs are not comparable because the latter used only the topic's short and often vague Query (Q) field,
while our runs use the Query and Description (Q+D) fields as the query.
We include the reference run uogTrA44xu to verify that our prime measures are similar to those of the track participants.
Note that non-prime metrics are much lower, due to the
low number of judged documents in the top twenty.
Our main result is over all 200 topics for years 2009 to 2012. 
\name\ significantly outperforms our BM25 baseline on both nDCG$'$@20 and MAP$'$
by a two-tailed paired $t$-test with Bonferroni correction for two tests at $p>0.05$. 
Compared to the Oracle Shard Model, which creates a single shard model for all of 
ClueWeb09, \name\ achieves 96\% of its nDCG$'$@20 performance.
This indicates that document scores from different shards are remarkably comparable,
leading to good hierarchical sharding effectiveness.
We conclude that \name\ is effective at scale,
outperforming a BM25 lexical baseline at ranking 
judged relevant documents and judged non-relevant documents.%

\begin{table*}[t]
  \caption{ClueWeb09 Results. $^\dagger$ indicates statistically significant improvement over BM25. \name\ percentages are percentage of Oracle.}
  \label{tab:clueweb09}
  \centering
  \begin{tabular}{lcc|ll|ll|c}
    \toprule   

Topics    & Queries & Approach & nDCG$'$@20 & MAP$'$ & nDCG@20 & MAP & Jg@20 \\
\midrule
Web12 & Q   & uogTrA44xu   & 0.348 &\bf{0.331}& \textbf{0.339} & \textbf{0.217} &  \textbf{1.000}  \\ 
Web12 & Q+D & BM25         & 0.355 & 0.245 & 0.070 & 0.049 & 0.194    \\
Web12 & Q+D & Oracle Model &\bf{0.457}& 0.296 & 0.113 & 0.060 & 0.160    \\
Web12 & Q+D & \name       & 0.437 & 0.288 & 0.100 & 0.058 & 0.152 \\
\midrule
Web09-12 & Q+D & BM25         & 0.323 & 0.217 & 0.094 & 0.058 & \textbf{0.300} \\
Web09-12 & Q+D & Oracle Model &\bf{0.448}$^\dagger$&\bf{0.278}$^\dagger$&\bf{0.132}&\bf{0.079}& 0.216    \\
Web09-12 & Q+D & \name       & 0.431$^\dagger$ (96\%) & 0.272$^\dagger$ (98\%) & 0.129 (98\%) & 0.077 (97\%) & 0.215 \\
   \bottomrule
\end{tabular}
\end{table*}

\begin{table*}[t]
  \caption{Neu{CLIR} MLIR Results. $^\dagger$ indicates statistically significant improvement over PSQ.}
  \label{tab:neuclir-mlir}
  \centering
  \begin{tabular}{ll|llll|c}
    \toprule   
Topics & Approach & nDCG@20 & MAP & R@100 & R@1000 & Jg@20 \\
\midrule
NeuCLIR22 & PSQ+HMM w/ Score Fusion & 0.315 & 0.195 & 0.269 & 0.594 & 0.901 \\
NeuCLIR22 & Oracle Model & 0.375 & \textbf{0.236} & \textbf{0.330} & \textbf{0.612} & 0.898 \\
NeuCLIR22 & \name\  & \textbf{0.381}$^\dagger$ & 0.228 (97\%) & 0.306 (93\%) & 0.602 (98\%) & \textbf{0.901} \\
\midrule 
NeuCLIR23 & PSQ+HMM w/ Score Fusion & 0.289 & 0.225 & 0.402 & 0.0.693 & 0.933 \\
NeuCLIR23 & Oracle Model & 0.330 & \textbf{0.281} & \textbf{0.468} & \textbf{0.760} & 0.922 \\
NeuCLIR23 & \name\  & \textbf{0.337}$^\dagger$ & 0.268 (95\%) & 0.431 (92\%) & 0.757 (100\%) & \textbf{0.927} \\

   \bottomrule
\end{tabular}
\end{table*}

Our second question is what data rate can be accommodated with a specific hardware configuration.
Here we compare \name\ to BM25;
the Oracle model is not considered because it does not represent the streaming setting that \name\ addresses.
Accounting for re-indexing as small shards are merged into large shards,
indexing the terabyte-scale ClueWeb09 collection requires about 71 days on an NVidia V100 GPU. 
The number of documents per shard was fixed as shown in Table~\ref{tab:collection-stats}; 
the number of passages varied from 24M to 51M for large shards with $A$ documents,
and from 2M to 7M for smaller shards with $B$ documents.
Large shards took on average seven hours to build,
while smaller shards required around thirty minutes. 
Thus we could index well over half a million documents per hour on this hardware.
Since shards can be searched in parallel on separate CPUs,
total search time is the search time of a large shard plus the time to merge results;
this averaged 1.3 seconds per query on ClueWeb09.
To field this configuration requires 108 CPUs each with 200GB of memory to hold the indexes loaded into memory and service the queries. %
The index requires 8.4 TB of disk.
For contrast, the sparse BM25 index used 0.4 TB of disk,
took 36 days of CPU time to build,
and BM25 query execution averaged 0.07 seconds on 1 CPU with 2GB memory.
Thus \name requires significantly more compute resources to build the index and to search.
Operationally, the added compute resources will need to be justified by the increase in effectiveness.

Our third question is about MLIR.
Table~\ref{tab:neuclir-mlir} compares \name\ to two other approaches on the NeuCLIR 2022 and 2023 datasets:
PSQ+HMM~\cite{darwish2003probabilistic, wang2012matching, xu2000cross, yang2024psq} with score fusion,
and an Oracle Shard Model. %
PSQ+HMM is a fast non-neural baseline
that relies on translation tables from statistical 
machine translation. This approach was chosen for its similarity to monolingual BM25 in being entirely non-neural. Since PSQ+HMM is a CLIR algorithm, MLIR is facilitated by 
score fusion across the languages without normalization. 
When \name\ ranks the top 1000 documents in each shard, it achieves essentially the same score as the oracle model for nDCG@20 on both query sets,
demonstrating the effectiveness of hierarchical sharding.
We hypothesize that the strong performance on nDCG@20 may be the result of two factors
that make it different from the ClueWeb experiments:
(1) there are fewer shards; and 
(2) NeuCLIR data is more recent and was crawled closer to its creation date,
so the dates associated with articles are likely more reliable than for ClueWeb.
On the NeuCLIR datasets,
Recall at 100 is most negatively affected by sharding relative to the oracle performance.
This is not seen in Recall at 1000,
so there may be differences arising from variations in the lossy vector compression
Finally, \name\ statistically outperforms the PSQ baseline as measured by nDCG@20
(two-tailed paired t-test with correction).

\section{Conclusion} %

\name\ provides a way to support multi-vector dense retrieval over a streaming collection of documents at large scale.
Its search speed supports interactive search. 
While the hierarchical sharding approach requires a document to be indexed more than once,
it allows for good performance on retrieval regardless of how long ago the document arrived.
While there is no doubt that utilizing a bi-encoder is more computationally expensive,
the multilingual pretrained language model has additional benefits over sparse retrieval approaches.
In a multilingual stream,
a document language might be unknown,
requiring the system to run Language Identification prior to PSQ+HMM;
this adds an additional source of error and performance degradation, which is eliminated for PLAID.
While \name\ may not be worth the computational expense for monolingual retrieval,
for multilingual retrieval PLAID, and thus \name, has clear advantage.%

\bibliographystyle{ACM-Reference-Format}
\bibliography{bibio}

\end{document}